# GPS Time Synchronization System for K2K[1]

H. G. Berns and R. J. Wilkes

Department of Physics, University of Washington, Seattle, WA 98195 USA

*Abstract*

The K2K (KEK E362) long-baseline neutrino oscillations experiment requires synchronization of clocks with ~100 nsec accuracy at the near and far detector sites (KEK and Super-Kamiokande, respectively), which are separated by 250 km. The Global Positioning System (GPS) provides a means for satisfying this requirement at very low cost. In addition to low-resolution time data (day of year, hour, minute, second), commercial GPS receivers output a 1 pulse per sec (1PPS) signal whose leading edge is synchronized with GPS seconds rollovers to well within the required accuracy. For each beam spill trigger at KEK, and each event trigger at Super-Kamiokande, 50 MHz free-running Local Time Clock (LTC) modules at each site provide fractional-second data with 20 nsec ticks. At each site, two GPS clocks run in parallel, providing hardware backup as well as data quality checks.

## I. Introduction

The K2K (KEK E362) long-baseline neutrino oscillations experiment (Fig. 1), which began data-taking in March, 1999, was designed to test atmospheric neutrino results from previous experiments indicating evidence for neutrino oscillations[1]. A newly constructed neutrino beamline at KEK, the Japanese national high energy physics laboratory located in Tsukuba, Ibaraki Prefecture, uses protons from the KEK 12 GeV proton synchrotron, targeted on a two-horn focusing system, to generate a broadband muon-neutrino beam of high purity (>98%) with mean energy on the order of 1 GeV. The secondary beam from the horn system passes through a 200m decay pipe, followed by a 100m earth berm, before entering the near detector experimental hall. The near detector system[2] consists of a 1 kT water Cherenkov detector followed by a Fine-Grained Detector system, composed of water tanks interleaved with 20 layers of scintillating fiber detectors, followed by a Pb-glass detector, and a muon detector. Super-Kamiokande[3], located 250 km away near Kamioka, Gifu Prefecture, serves as the far detector. Super-Kamiokande is located in a mine with >1 km rock overburden in all directions.

Trigger rates at Super-Kamiokande are low enough that the expected arrival time window for KEK beam neutrinos only needs to be determined to within a few microsec. The Global Positioning System (GPS) provides a means for easily satisfying this requirement at very low cost. This paper will decribe the time synchronization system constructed for that purpose, which in fact has accuracy on the order of 100 nsec.

## II. GPS Timing

GPS consists of 27 satellites maintained by the US Department of Defense (DOD), each transmitting coordinated "GPS Time" according to its onboard atomic clock[4]. GPS Time differs from UTC only in the absence of the leap seconds which are periodically inserted in UTC. Most GPS receivers (including ours) automatically take the shift into account using data downloaded from the satellites, so the time reported is UTC. The satellites' onboard clocks are regularly conditioned to match GPS time according to a ground-based reference clock system (actually a large number of high-precision atomic time standards). The satellites also broadcast their ephemerides, so their position in space can be accurately calculated as a function of time. The ephemerides also are regularly recalculated and updated. With each satellite's position in space known to high accuracy from its ephemeris, users' receivers can fit for their position and time (x,y,z,t) if four or more satellites are simultaneously in view.

Since the GPS satellites are constantly referenced to a national standards laboratory time base, the GPS system provides a simple and inexpensive way to obtain high precision absolute time, synchronized to UTC, without purchasing and constantly recalibrating a set of atomic clocks. The GPS system is designed to give standard errors of about 150m on a single position fit and 150 nsec relative to UTC on a single time fit. For a fixed antenna location, as in K2K, long-term averaging of the measured antenna coordinates provides improved accuracy.

## III. Overview of the K2K Time Synchronization System

The GPS system provides UTC (Universal Time, Coordinated) timestamps for K2K data at both sites. Beam spill times are logged at KEK, and each event trigger is timestamped at Super-Kamiokande (SK). An overall block diagram for the K2K system is shown in Fig. 2, and for the SK system in Fig. 3.

The system hardware consists of the following main components at each site:

- Primary receiver (TrueTime Model XL-DC [5])
- Backup receiver (Motorola UT+Oncore [6])
- Local Time Clock (LTC) board

The LTC board, designed and built at UW, is a VME 6U board containing a free-running 50 MHz oscillator/counter, with additional circuitry required for interfacing.

At KEK, the backup receiver and the LTC board are housed in a VME crate located in the North Hall Control Room, which is connected by a Bit-3 interface to a Pentium

---

[1] Work supported by US-DOE grant DE-FG03-96ER40956/B.

PC running Slackware Linux. The primary receiver is mounted in the same rack, and the antennae are mounted on the control room roof.

At Super-K, the LTC board sits in a VME crate in the detector's central electronics hut, connected via Bit-3 interface to a Sun Sparc-20 running Solaris. The antennae are mounted on an external building near the mine entrance, which is also used to house the receivers, which are connected to the central hut via a 2 km optical fiber with electro-optical converters at each end. Delay time introduced by the optical fiber link has been directly calibrated.

Each receiver produces two kinds of output: a 1-PPS square-wave signal whose leading edge is correlated with the beginning of UTC seconds to within the specified precision, and an ascii data stream containing complete GPS data (latitude, longitude, altitude, date, time down to milliseconds, housekeeping data). The 1 PPS signal is used to calibrate the LTC counter. At each 1-PPS leading edge, the LTC counter reading is recorded. Thus the actual number of nominal 20 nsec LTC oscillator "ticks" per UTC second is calibrated. A 300-sec running average for the oscillator rate is maintained.

Upon receipt of each each trigger pulse (representing time of beam spill at KEK, or hardware trigger at Super-K), the LTC count and the GPS ascii data are latched and recorded. The LTC count provides the fractional second of the trigger, down to 20 nsec precision, accurately synchronized with UTC within 100 nsec. The ascii data provide the date and coarse time down to seconds. At KEK, the TrueTime receiver has an additional option module installed which records the "event time" (trigger pulse arrival) directly, with 30 nsec precision. Thus we have redundant estimates of the spill time in high resolution, from the commercial receiver plus our own LTC.

Fig. 4 shows how high-precision trigger times in UTC are derived from the 1-PPS and LTC data. All raw data are logged at both sites for all triggers, to provide a backup for the realtime data transmission between sites, and also to provide information for continuous data-quality checking.

## IV. SYSTEM PERFORMANCE

The KEK system was set up at the end of September, 1998. In October, 1998, the SK system, which had been in operation since 1996 with only the TrueTime primary clock and an earlier version of the LTC board, was upgraded to match the KEK system.

GPS receivers require 4 or more satellites in view simultaneously to initially determine their antenna's geodetic coordinates and time shift relative to GPS time. After this survey, which typically takes 24 hr, time synchronization can be maintained with only one satellite in view at a time. For a survey covering several days, we had 4 or more satellites in view 100% of the time at KEK and 1 or more in view all of the time at Super-K. The probability of having only 1 satellite in view is about 6e-05 at Superk, and zero for KEK. During brief periods of satellite blackout, time synchronization depends upon internal oscillator stability. The TrueTime receiver provides drift less than 1 part in $10^6$ per 24 hr, so the system is extremely unlikely to lose synchronization during very brief blackouts.

Fig. 5 shows the time jitter of the 1 PPS leading edges of the KEK Motorola clock, relative to the TrueTime 1 PPS signal. The jitter distribution has HWHM <100 nsec.

Fig. 6 shows LTC oscillator drift statistics. Although the LTC uses a simple, uncompensated 50 MHz quartz oscillator circuit, the technique of 300-sec running averages provides more than adequate effective stability, allowing overall system performance well within the limits required. The histogram indicates that deviations >1 Hz/50 MHz are unlikely, and deviations greater than 3 Hz/50 MHz are not seen.

## V. ACKNOWLEDGEMENTS

We gratefully acknowledge comments, suggestions and assistance from many members of the K2K and Super-Kamiokande collaborations. The authors alone are responsible for any errors or omissions in this paper.

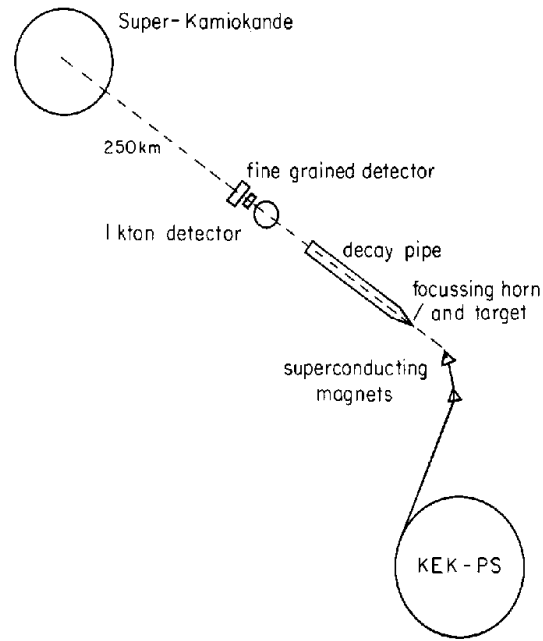

Figure 1: K2K (KEK E362) long-baseline neutrino experiment.

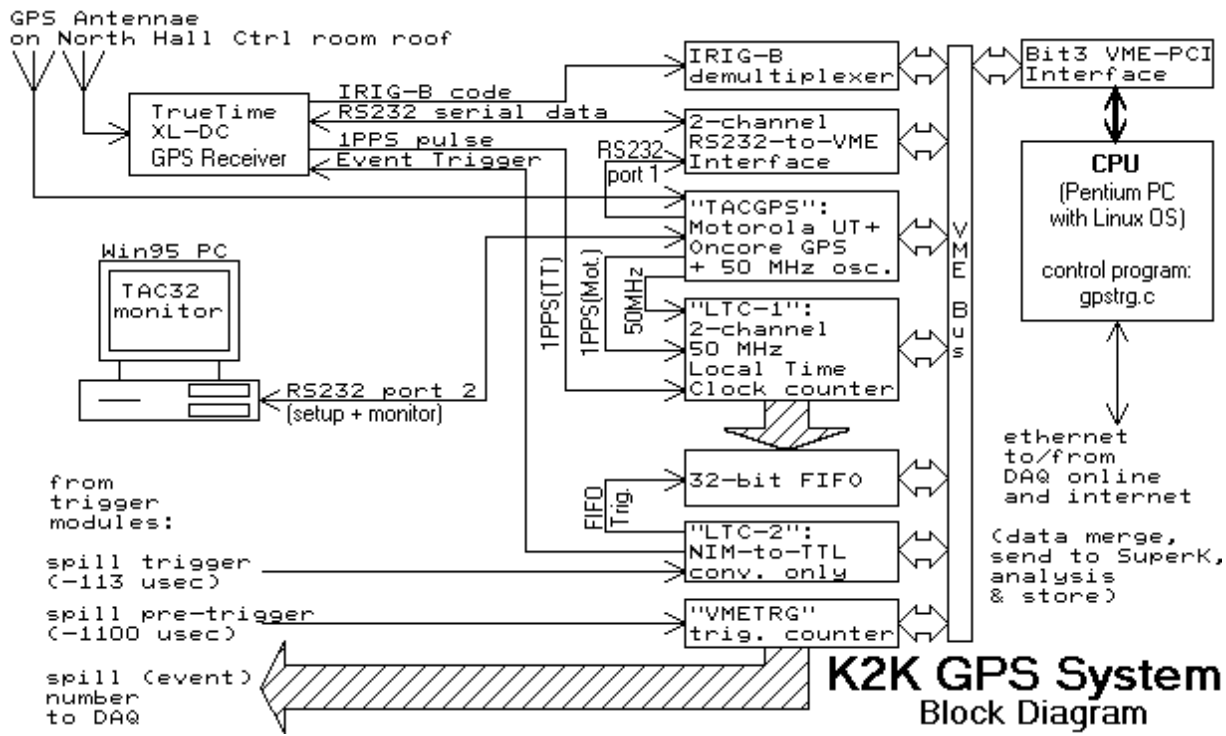

Figure 2: Block diagram of the K2K GPS system at KEK location.

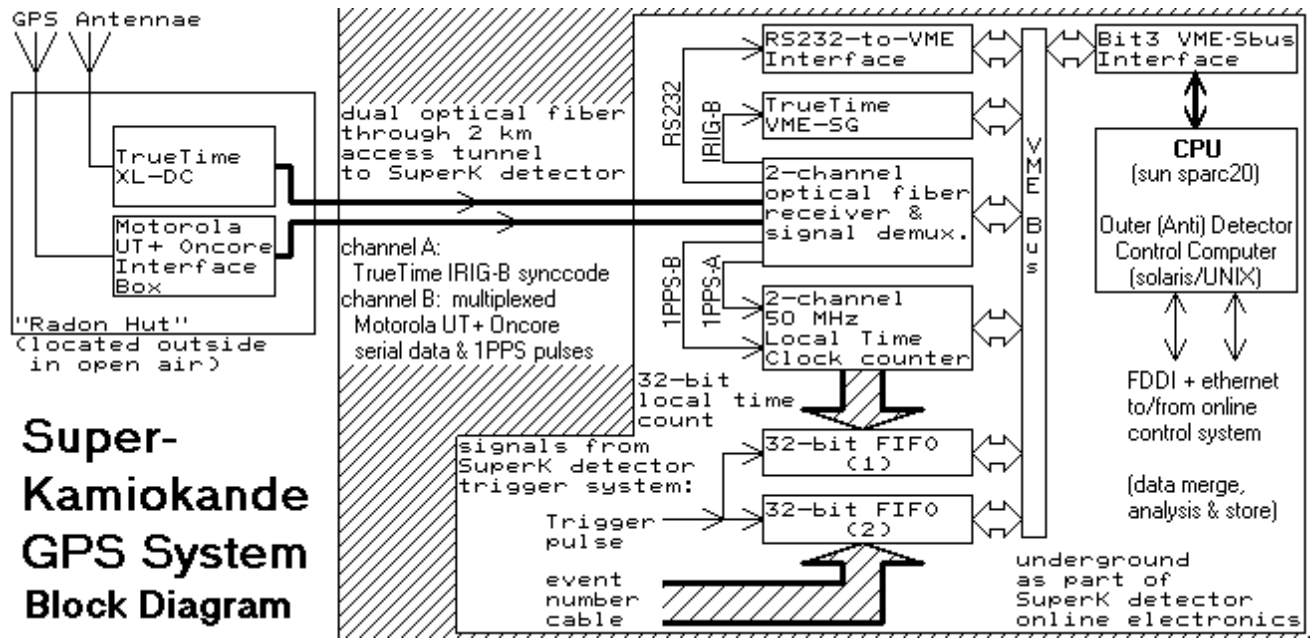

Figure 3: Block diagram of the Super-Kamiokande GPS system.

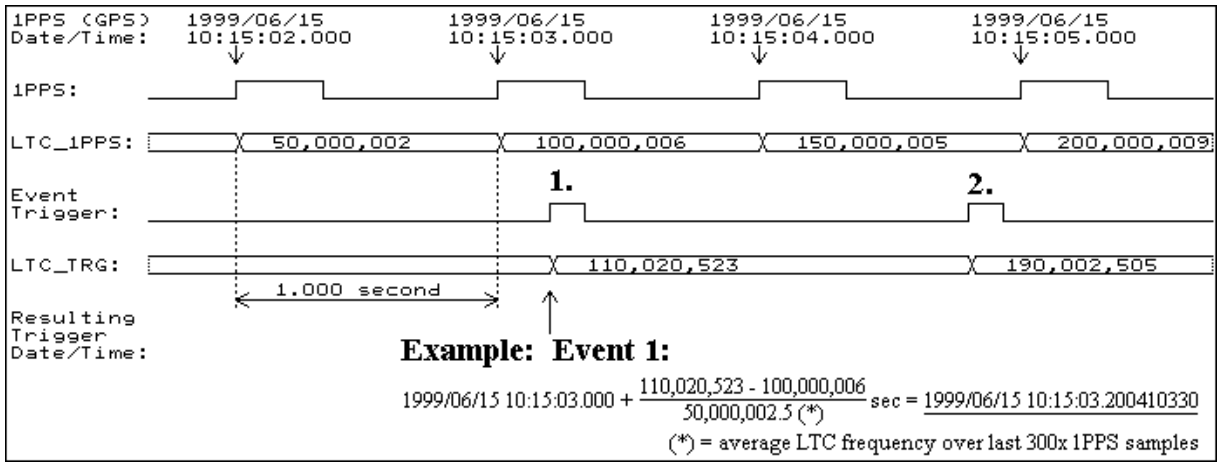

Figure 4: 1-PPS and trigger pulse timing diagram with example for computing high-resolution trigger time.

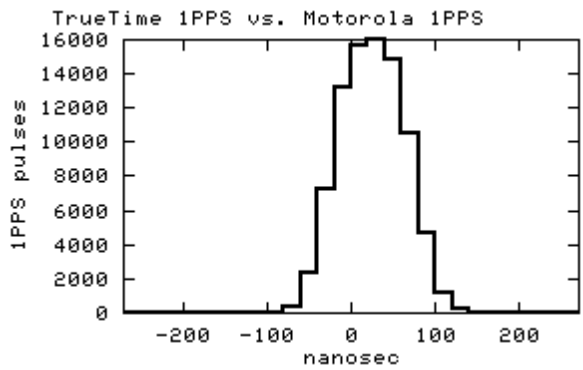

Figure 5: Histogram of jitter between the TrueTime and Motorola 1PPS pulses, direct comparison over 24-hour period.

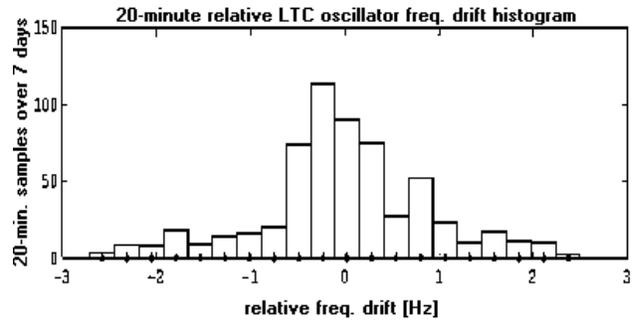

Figure 6: Relative drift in oscillator frequency (50 MHz nominal) between 20-minute samples over 7-day period.